# UPDATE FOR "THE COLLAPSE OF BELL DETERMINISM"

James D. Malley

*Center for Information Technology, NIH, Bethesda MD, USA*

**ABSTRACT**   We update our paper "The collapse of Bell determinism" (*Physics Letters A*, 359 (2006): 122-125; available online 16 June 2006). First, we point out that Olivier Brunet, using lattice theoretic methods, has recently, and quite independently, derived the core technical lemma of our paper; see "A priori knowledge and the Kochen-Specker theorem," (*Physics Letters A*, available online 5 January 2007). He has also kindly pointed out a misstep in the last line of one of our results. We discuss the correction and comment on slightly revised notions of how Bell determinism collapses. We also correct a few typos in the original paper.

**ALTERNATIVE DERIVATION FOR KEY LEMMA**   In [1] we began with the usual conditions for Bell determinism, namely that there exists an assignment of 0 or 1 to every vector $x$ in a state space $H$ such that the assignment, $v(x)$, respects orthogonality. We then used elementary geometry to derive our central technical result, namely

*Lemma 3*. Suppose $v(g) = 1,$ for unit vector $g$. Then for any vector $y \in H$ not in the span of $g$ we must have $v(y) = 0$.

The corresponding result in Brunet [2] is his *Theorem 7*; see his comment following *Proposition 8*. It was interesting to learn that his derivation was completed independently, and just a few months after our own paper [1]. Moreover, he used an alternative technology, that of Sasaki filters and lattice theoretic methods, along with a bit of geometry.

**EXACTLY WHAT IS COLLAPSING?**   In a subsequent email conversation Brunet pointed out a faulty last step in our proof of *Theorem 1*. Specifically, in the proof we first show that given any vector $x$ such that $v(x) = 1,$ then any other vector $h$ is either in the span of $x$ or orthogonal to $x$. This argument is valid in $\dim H \geq 3$. However the last step in the proof asserts that what was true for arbitrary $h$ must be true as well for any other vector $t$, not in the span of $x$ or orthogonal to $x$. In particular, for $t = h + x,$ it would seem to follow that $t \perp x$ or $t$ in the span of $x$, which implies that $h$ must be in the span of $x$. Brunet saw correctly that the conclusion for vector $t$ is problematic since the conclusion for $h$ (assuming the Bell determinism conditions are valid) already conflicts with the structure of the state space $H$. Thus, $H$ can't even be a vector space if arbitrary $h \in H$ is forced to be either in the span of $x$ or orthogonal to $x$. Using the notation *BKS* for the Bell determinism conditions, as given in [1], it is therefore important to restate *Theorem 1* in [1] as

*Theorem 1 (revised).* Given the *BKS* conditions suppose $x \in H$ is such that $v(x) = 1$.

Let $h$ be any vector in $H$. Then $P_x = P_h$ or $P_x \perp P_h$.

Of course one summary conclusion here is still that Bell determinism is not valid for any state space $H$, $\dim H \geq 3$. But we can refine this conclusion by reframing the "collapse" of Bell determinism of [1]. Here is one version: Bell determinism for a state space $H$, $\dim H \geq 3$, violates the vector space structure of the ambient state space $H$. And here is another, more subdued version: In any state space $H$, let $\{x_i\}$ be an orthogonal basis and let $y = \Sigma\, x_i$; then under Bell determinism $y$ must be equal or orthogonal to one of the $x_i$.

We remark that even the quieter version drives home the central consequence of the Bell determinism conditions. Thus it is not simply that we can, with sufficient craft and diligence, locate some small, distinguished set of vectors that violates the conditions. It is rather that with an alternative, tiny geometric assemblage it follows that *every* basis set for the ambient space $H$ generates a violation of the conditions.

**WHAT IS THE SOURCE OF THE COLLAPSE?** Our result, and that of Brunet [2] do not exist in isolation, so we provide some context. We already know that the Bell determinism conditions, coupled with the full power of Gleason's theorem, lead at once to a sharp conflict. More directly, we can use the real-valued Hilbert space version of Gleason's theorem as derived by Gudder [3; *Corollary 5.17*]. Specifically, suppose that all one-dimensional projectors are assigned a value 0 or 1, and that the assignment respects orthogonality. If some distinguished projector, $P_x$, is known to be assigned the value 1, then, by *Corollary 5.17* we must have for any vector *y*, that $v(y) = v(P_y) = tr[DP_y] = |\langle x, y \rangle|^2$. But then our collapse result (any version) is immediate, since by assumption $v(y) = 0$ or 1.

However, Gleason has always seemed much too powerful and thus somewhat uninformative for studying the problem of Bell determinism. Hence, we suppose, the generation of the long historical arc composed of the many elegant proofs with small sets of vectors. These certainly have the virtue of simplicity, but to us they have seemed rather mysterious.

There is an intermediate path here that is possibly more informative. Thus, as noted above, Gudder has shown how to derive Gleason (over real Hilbert spaces, *H*, with $\dim H \geq 3$) using a somewhat long sequence of basically elementary, but clever geometric steps. This derivation also requires a continuity argument at several points, and a few other analytic facts, e.g., on a bounded set of reals, a monotone function can have at most a countable number of discontinuities. However, almost all of the many careful steps in Gudder's derivation are not required when it is given that the valuation makes *only* assignments of 0 or 1 to projectors, and such is the case with Bell determinism. The collapse then can be seen as the collision of this form of determinism with something not specifically quantum but rather with simple features of real inner product spaces.

**SOME HOUSEKEEPING** We conclude by correcting some typos in [1]:

(*1*) p. 123, right column, line 13: replace $P_Z$ with $P_Y$

(*2*) p. 123, right column, line 30: replace $\langle y, y \rangle_S$ with $\langle y, y \rangle$

(*3*) p. 124, right column, line 24: replace $\langle x, y \rangle_S$ with $\langle x, y \rangle$

**REFERENCES**

[1] JD Malley. The collapse of Bell determinism. *Physics Letters A* 359 (2006): 122-125. Available online 16 June 2006.

[2] O Brunet. A priori knowledge and the Kochen-Specker theorem. *Physics Letters A* (2007). Available online 5 January 2007.

[3] S Gudder. **Stochastic Methods in Quantum Mechanics**. North-Holland. 1979.